\documentclass[a4paper, reprint, twoside]{revtex4-2}

\usepackage[english]{babel}
\usepackage[utf8]{inputenc}
\usepackage[labelfont=up]{subcaption}
\usepackage[x11names]{xcolor}
\usepackage{float}
\usepackage{mathtools}
\usepackage[top=1.in, bottom=1.in, left=0.5in, right=0.5in]{geometry}
\usepackage{amsthm, amssymb}
\usepackage{graphicx}
\usepackage{lipsum}
\usepackage{bm}
\usepackage[pdftex, pdftitle={Article}, pdfauthor={Author}]{hyperref}
\usepackage{wasysym}
\usepackage{enumitem}
\hypersetup{colorlinks=true, citecolor=black, linkcolor=black, urlcolor=blue}

\begin{document}

\title{Emergence of Scale-Free Networks in Social Interactions among Large Language Models}
\author{Giordano De Marzo$^{1, 2, 3, 4}$, Luciano Pietronero$^{1}$ and David Garcia$^{2, 5}$}
\affiliation{
$^1$Centro Ricerche Enrico Fermi, Piazza del Viminale, 1, I-00184 Rome, Italy.\\
$^2$Complexity Science Hub Vienna, Josefstaedter Strasse 39, 1080, Vienna, Austria.\\
$^3$Dipartimento di Fisica Universit\`a ``Sapienza”, P.le A. Moro, 2, I-00185 Rome, Italy.\\
$^4$Sapienza School for Advanced Studies, ``Sapienza'', P.le A. Moro, 2, I-00185 Rome, Italy.\\
$^5$University of Konstanz, Universit\"atstra{\ss}e 10, 78457 Konstanz, Germany}

\date{\today} % Leave empty to omit a date

\begin{abstract}
    Scale-free networks are one of the most famous examples of emergent behavior and are ubiquitous in social systems, especially online social media in which users can follow each other. By analyzing the interactions of multiple generative agents using GPT3.5-turbo as a language model, we demonstrate their ability to not only mimic individual human linguistic behavior but also exhibit collective phenomena intrinsic to human societies, in particular the emergence of scale-free networks. We discovered that this process is disrupted by a skewed token prior distribution of GPT3.5-turbo, which can lead to networks with extreme centralization as a kind of alignment. We show how renaming agents removes these token priors and allows the model to generate a range of networks from random networks to more realistic scale-free networks.
\end{abstract}

\maketitle
\section{Introduction}
    \noindent
    The integration of Artificial Intelligence (AI) into our daily lives has been subtle yet profound, especially with the advent of Deep Learning technologies. This quiet infiltration has taken a dramatic turn with the rise of Generative AI (GenAI) and, more specifically, Large Language Models (LLMs) like ChatGPT. These technologies, once cloaked in the guise of research assistance tools, have now emerged at the forefront of innovation, becoming both instruments and subjects of scientific inquiry. They have transcended their initial roles in programming and writing assistance, enabling their use for tasks they were not designed nor trained for \cite{bubeck2023sparks}. In this context, a significant amount of attention has been devoted to understanding and analyzing the capabilities of individual models \cite{chang2023survey, xu2022systematic, valmeekam2022large}. These studies have provided invaluable insights into the nuanced functionalities, biases, and potential applications of LLMs \cite{argyle2023leveraging, jansen2023employing, park2022social, thirunavukarasu2023large, abdulhai2023moral, singhal2023large, frank2023baby, webb2023emergent}. However, an area that warrants greater exploration is the interaction between multiple LLMs \cite{betz2022natural,tornberg2023simulating, grossmann2023ai}. This aspect of research is crucial for several reasons, especially when considering how human social interactions can lead to emergent collective behaviors \cite{garcia2023psychology}.

    The study of the behavior of LLMs in isolation, while informative, offers a limited perspective on the potential dynamics of these models in more complex, interactive settings. In human societies, many of the most significant processes and phenomena are not the result of individual actions but are collective outcomes. For instance, cultural trends, economic shifts, and social movements are all emergent properties of numerous individuals interacting with each other, often in complex and unforeseeable ways \cite{anderson1972more}. LLMs are an integral part of the concept of generative agents: modeling human behavior with LLMs in combination with other structures and properties such as memory, plans, and identities \cite{park2023generative,zhang2023building}. When multiple generative agents interact in a simulation, they may exhibit behaviors and generate outcomes that are not predictable or directly coded into any single agent or LLM \cite{he2023homophily}. Instead, they arise from the complex interactions and interdependencies between multiple agents and their LLMs. In a way, each generative agent contributes to a collective 'intelligence' or pattern that is more sophisticated and unpredictable than its individual capabilities would suggest. For instance, simulations with generative agents show examples of emergent behaviors like information diffusion and coordination \cite{park2023generative}. 

    Complex networks are an emblematic example of emergent structures \cite{capocci2006preferential}. Complex networks have scale-free degree distributions with surprising emergent properties: the variance of degrees can grow with the size of the network and epidemic spreading can be extremely hard to tackle \cite{pastor2001epidemic}.
    For instance, the World Wide Web and online social networks are formed from the interactions of countless individuals, where relationships and information flows create a dynamic, evolving structure. In particular, online social networks with follower links, such as Twitter or Instagram, have been shown to have scale-free distributions \cite{teng2015toward, aparicio2015model}.    
    These examples highlight a key feature of complex networks: their structure and dynamics are an emergent product of the interaction between their components, rather than being dictated by a central entity controlling the system. 
    
    In this article, we aim to a first understanding of how generative agents can self-organize into network structures, and whether these can be complex. Just as human interactions give rise to the complex structures of social networks, we hypothesize that interactions among generative agents can lead to the formation of similarly intricate networks. The significance of this research is underscored by two emerging trends: on one hand, the rise of AI-driven social networks like Chirper, and the anticipated increase in LLM-governed bots within conventional social networks \cite{luo2023analyzing, he2023homophily}; on the other hand, the development of Agent-Based Models using generative agents \cite{grossmann2023ai, tornberg2023simulating}. These developments signal a shift towards a digital ecosystem shaped by LLMs, where interactions between AI entities could become as commonplace as human communications.

\section{Results}
    \subsection{Simulating Social Networks Growth}
    
        We model a simplified version of a growing online social network inspired by complex network growth models, such as the Barabasi-Albert model \cite{barabasi1999emergence}, where nodes enter the network sequentially and connect to other nodes. In our simulations, nodes are generative agents that represent the users of an online social network. Instead of explicitly modelling and simulating the actions of agents with an equation, we utilize an LLM to guide the actions of the generative agents. 
        
        \begin{figure}
%            \centering
            \includegraphics[width=0.5\textwidth]{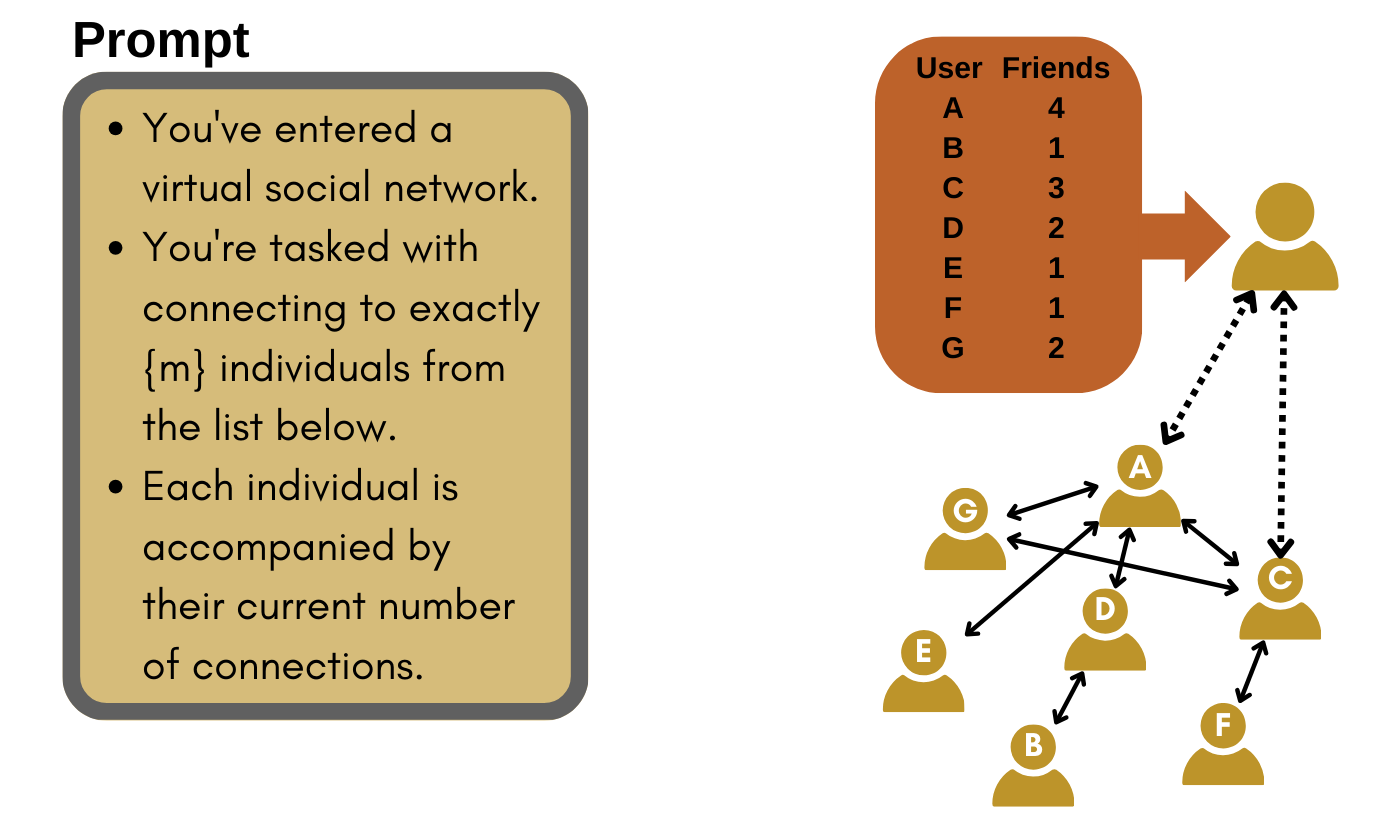}
            \caption{\textbf{LLM-based Network Growth with generative agents.} Left: The prompt we used to initialize the LLM of each generative agent as a user of an online social network. This prompt is followed by a comprehensive list of all network users along with their numbers of friends. Right: A schematic representation of the simulation process. At each timestep, a new agent is instantiated using the given prompt and receives the degree information of existing network users. The agent then identifies $m$ users it wishes to connect with and the network is updated to include these new connections. This iterative process continues with the addition of new agents to the network.}
            \label{fig:fig1}
        \end{figure}

        Fig.~\ref{fig:fig1} illustrates how our agents behave in the simulation. At each timestep and using the OpenAI API, a generative agent powered by GPT-3.5-turbo is initialized using the prompt shown in Fig.~\ref{fig:fig1} (left), representing a user in an online social network tasked with selecting $m$ users to connect with. The agent is presented with a list of all existing other agents, each identified by a random 3-character string, along with their network degree, i.e., the number of connections they already possess, as illustrated in Fig.~\ref{fig:fig1} (right). The order by which users are listed is completely random and changes at each time step to avoid potential biases due to the placement in the prompt. The new agent then autonomously decides which other agents to connect with by replying with their names. 
        Note that the information agents receive in our simulations is identical to the information that nodes have in the Barabasi-Albert model. The main difference lies in the dynamics of creating connections: while nodes in the Barabasi-Albert model follow a rigid program in which the agent connects given a fixed probability rule, our agents generate the list of nodes to connect to based on the text generated by the LLM. Once the generative agent selects the users, the corresponding $m$ undirected links are formed, and the network's degree list is updated. This process is then iteratively repeated with the addition of new nodes to the network until reaching the desired network size.
    
    \subsection{Network Topologies}
        \begin{figure*}
            \centering
            \includegraphics[width=0.99\textwidth]{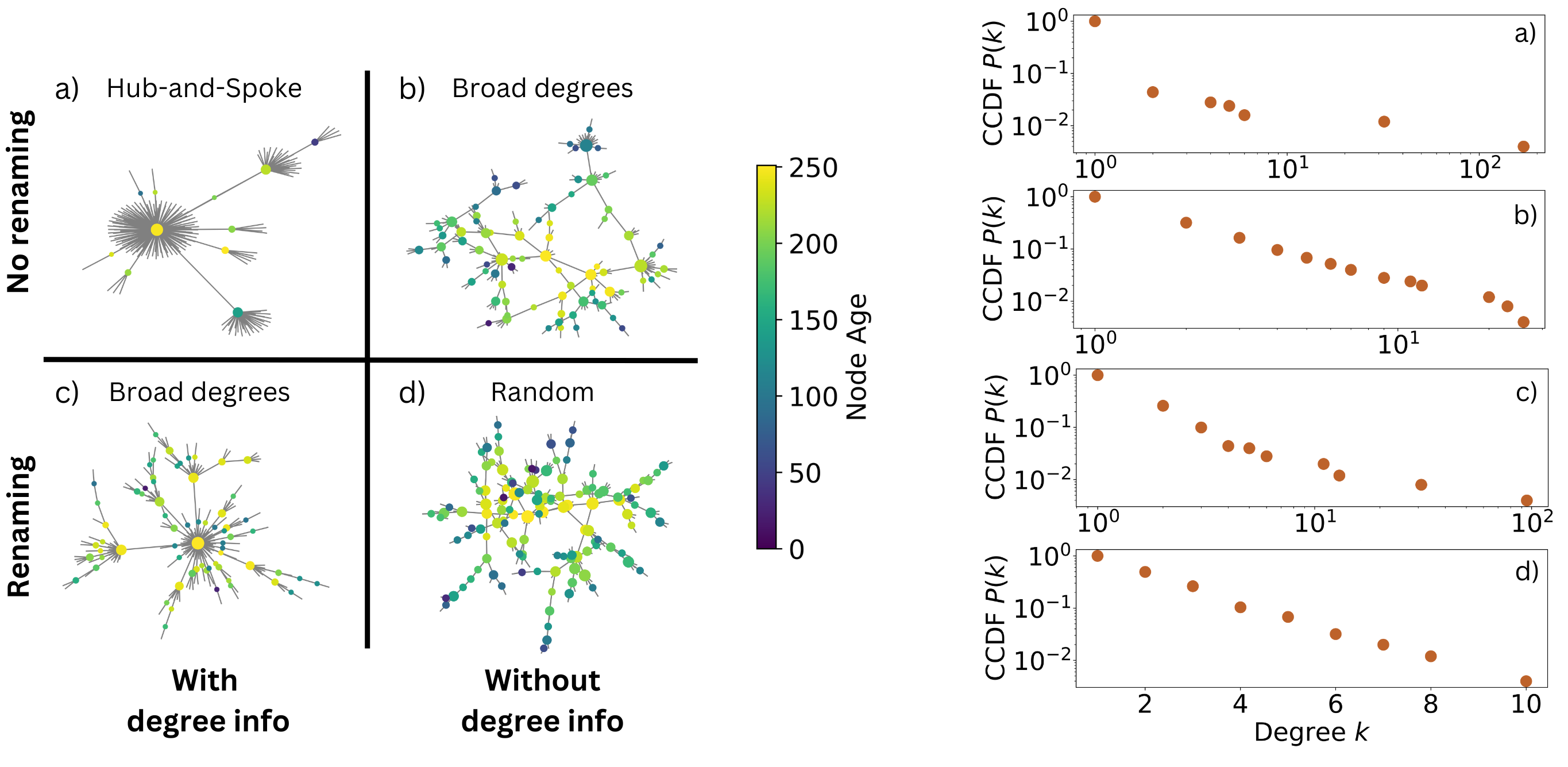}
            \caption{\textbf{Network Topologies.} Left: Different network topologies corresponding to the four scenarios discussed in the main text: presence/absence of token prior and presence/absence of degrees in the agent interface. Colors show the age of nodes, while the node size is proportional to the logarithm of the degree. Right: Complementary Cumulative Distribution Functions (CCDF) for the four different network topologies.}
            \label{fig:fig2}
        \end{figure*}

Upon establishing our network growth process, we focus on analyzing the resulting network structures. When agents can see the number of friends of others, the resulting network has a hub-and-spoke topology where very few agents garnered the extreme majority of links, as shown on the left panel of Fig.~2,a. In this case, incoming connections from new nodes are restricted to just a handful of previous nodes, leading to an inequality in degrees that is larger than what can be observed in empirical social network structures. This extreme concentration can be seen in the CCDF on the right panel of Fig.~2. This is a kind of Artificial Intelligence alignment that leads to unrealistic network structures and extreme concentration in the degrees of the network.

To better understand this network generation process, we changed the model to remove the degree information shown to the agents, i.e. the number of friends of each node is not shown on the interface that agents see, they only receive the list of names of other agents to select $m$ from. By analogy to the Barabasi-Albert model, one would expect that this generates a random network structure. However, this simulation without degree information for the agents surprisingly leads to a network with a broader degree distribution that topologically resembles better empirical online social networks than the case in which degrees were shown, as can be seen on the left and right panels of Fig.~2,b. A closer inspection of this network topology shows its main problem: node age is not correlated with node degree and new nodes might be able to attract a disproportionately large degree, which is not the case in network growth models with preferential attachment. This unexpected outcome can be linked to the presence of a non-trivial prior in the token generation process of the LLM. Each agent in our network is assigned a random name, and the LLM exhibits a preference for certain letter sequences that might be biased by their frequency in the training corpora of the LLM. This could be a result of Zipf's law when using letters or Benford's law when using numbers \cite{zipf2016human, pietronero2001explaining, de2021dynamical}. This preference leads to increased popularity and connectivity for certain agents just because they happen to have a name with a more frequent prior in the LLM. Therefore, the broad distribution we observe is a consequence of inherent biases in the responses of GPT3.5-turbo, rather than a result of preferential attachment being reproduced by the generative agents in our simulation.

To validate our suspicion about the prior probabilities of GPT3.5-turbo influencing the outcome, we adjusted the simulation by randomly renaming agents in each iteration, simply permuting their names but keeping their references in the network and this way also keeping track of their degrees. This version of the model with renaming, in combination with node degree information in the interface of agents, generates a complex network structure similar to empirical networks as shown on the left panel of Fig.~2,c. When inspecting the age of nodes, this network displays the typical correlation between node age and degree, resembling better what is produced by a preferential attachment process. If we change the interface to hide the node degree information, while keeping the renaming, the result resembles a random network, as shown on the left panel of Fig.~2,d, with a narrow degree distribution shown on the right panel (notice the linear horizontal axis). This way, after renaming agents in the interface, the model ranges from a complex network structure when agents have degree information to a random network structure when they do not have access to the degree of others. Building on this, the Hub-and-Spoke structure of the model without renaming and with degree information is a combination of two concentration processes: a kind of preferential attachment and the prior of token names, which leads to an extremely centralized network.
        
\subsection{Scale Free Networks}

As shown above, when the token prior is eliminated by renaming, we observe a shift from a random network to a more complex structure when agents are able to see the number of friends of other agents. When no renaming is done, a hub-and-spoke topology emerges when agents can see the number of friends of other agents. These cases can be understood when considering the different regimes of the Barabasi-Albert model \cite{barabasi1999emergence}, where the likelihood of establishing a connection to a particular node is proportional to its degree \( k \), expressed through the linking probability \( \pi(k) \), given by:
        \begin{equation}
            \pi(k) \sim k^{\mu}.
            \label{eq:linking_probability}
        \end{equation}
        This mechanism, known as preferential attachment, implies that nodes with higher degrees have a greater chance of acquiring new links. The attachment exponent \( \mu \) plays a crucial role in shaping the network's topology \cite{dorogovtsev2000structure}:
        \begin{itemize}
            \item For \( \mu = 0 \), the network is random with an exponential degree distribution.
            \item With \( 0 < \mu < 1 \), the network exhibits a stretched exponential degree distribution.
            \item At \( \mu = 1 \), the network becomes scale-free, characterized by a power-law degree distribution.
            \item For \( \mu > 1 \), the network evolves into a hub-and-spoke structure, often described as a 'rich-get-all' scenario.
        \end{itemize}

Typical online social networks have degree distributions that correspond to linear preferential attachment, where values deviating from this narrow range result in distributions that are markedly different.

 \begin{figure*}
            \centering
            \includegraphics[width=0.9\textwidth]{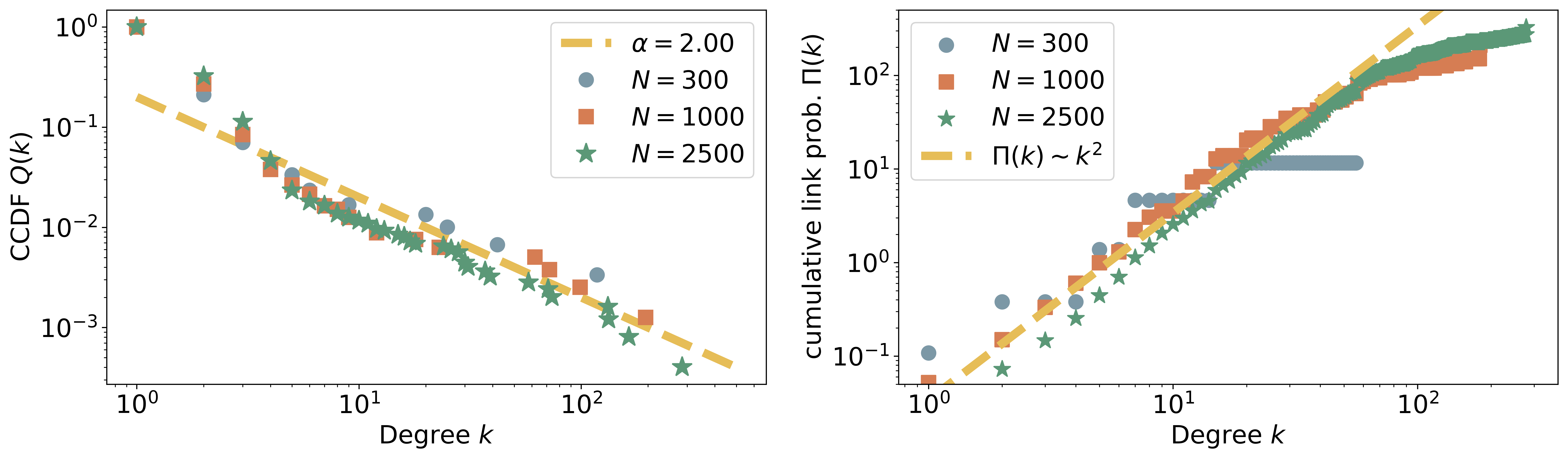}
            \caption{\textbf{Linear Preferential Attachment.} Left: CCDF of the degrees for three network sizes $N=300, 1000, 2500$. The solid line is a guide for the eyes and corresponds to a power law PDF with exponent $\alpha=2$. The network exhibits a scale-free structure. Right: Reconstructed cumulative linking probability $Pi(k)$ as function of $k$ for the three network snapshots. As $N$ increases, a longer quadratic regime appears. This corresponds to a linear preferential attachment, typical of scale-free networks and human interactions.}
            \label{fig:fig3}
        \end{figure*}

Our model with generative agents mirrors the regimes seen in the Barabasi-Albert model, where the degree distribution evolves from an exponential to broader probabilities as the preferential attachment exponent $\mu$ increases above zero, eventually leading to a rich-get-all all scenario for $\mu>1$. Our aim is to ascertain the regime (stretched exponential random network, scale-free network, hub-and-spoke network) to which the networks produced by our generative agents belong. Fig.~\ref{fig:fig3} (left) shows the evolution of the Complementary Cumulative Distribution Function (CCDF) of node degrees as the network grows to $N=2000$ nodes. The CCDF's linear trend in a log-log plot indicates a scale-free topology. The solid line in the figure represents the Maximum Likelihood (ML) fit to the CCDF \cite{clauset2009power, alstott2014powerlaw}, corresponding to a power-law degree distribution with exponent $\alpha=1.93\pm0.12$. %Additionally, we report the distribution of the power-law exponent, calculated over $20$ simulations with $N=500$ nodes each.  
        %Detailed methodology for the fitting procedure and goodness-of-fit tests can be found in the Methods section.
        
        Our findings suggest that generative agents, akin to humans in online interaction, tend to spontaneously form scale-free networks. However, due to the relatively limited size of the network and the costs associated with simulating larger networks, further in-depth analysis is required to confirm that generative agents display linear preferential attachment. We reconstruct the attachment probability $\pi(k)$; where a linear relationship with the degree $k$ in this probability would indicate a scale-free network governed by linear preferential attachment. Following the methodology outlined in \cite{newman2001clustering}, we examine successive network snapshots to calculate the likelihood of new nodes connecting to existing nodes with specific degrees, thus deriving $\pi(k)$. To minimize noise, we compute the cumulative attachment probability $\Pi(k)=\sum_i^k\pi(k)$ and plot it against $k$. This cumulative metric exhibits quadratic growth in the presence of linear preferential attachment. Fig.~\ref{fig:fig3} (right) shows $\Pi(k)$ for the three network sizes shown above. The plots reveal an increasingly pronounced quadratic shape as $N$ grows, indicating linear preferential attachment among generative agents. 

\section{Discussion}
    Large Language Models (LLMs) have demonstrated their proficiency in mimicking human language, achieving a level of text generation that often becomes indistinguishable from that authored by humans. This capability has sparked interest in leveraging LLMs for a new generation of Agent-Based Models (ABMs), promising an unprecedented level of realism \cite{grossmann2023ai} in which agents can even serve as a kind of \emph{digital twin} of a social media user. Such advancements hold immense potential for applications ranging from testing social platform interventions \cite{tornberg2023simulating} to analyzing epidemic spread \cite{williams2023epidemic}. However, realizing this potential hinges on understanding the extent to which LLMs can not only replicate human behavior on an individual level but also exhibit collective behaviors that are fundamental to human societies. Since most LLM benchmarks are structured around question-answering, there's limited insight into LLMs' abilities to exhibit group behaviors that can only be observed in collectives where many individuals interact with each other and their environment.
    
   In this study, we explored this question by focusing on a fundamental emergent pattern in human interactions: scale-free networks. These networks, prevalent across various systems from biological to socio-economic domains, are typified by a topology where degree distributions follow a power law, and thus most nodes hold few connections while a minority are highly connected.
   %The Barabasi-Albert model and the concept of preferential attachment are pivotal in explaining the evolution of these networks through simple individual interactions. Systems such as the World Wide Web and co-authorship networks have been demonstrated to follow linear preferential attachment, suggesting a universal human propensity to linearly favor popularity \cite{redner2005citation, barabasi1999emergence}. 
   Our findings reveal that a model with generative agents driven by GPT3.5-Turbo, the LLM that ChatGPT uses, parallels human behavior in online social networks with respect to their scale-free degree distribution. Networks with generative agents exhibit a similar tendency for linear preferential attachment, resulting in the formation of scale-free networks. Notably, the scale-free networks generated in our simulations exhibit a power law exponent close to $-2$, the famous Zipf's law \cite{zipf2016human, de2021dynamical}, a value very close to that observed in online social networks such as Twitter \cite{aparicio2015model}.
   
    Conclusively, our results underscore the challenges in incorporating LLMs as generative agents. Often described as ``stochastic parrot'' \cite{bender2021dangers}, LLMs are generally understood as systems that randomly generate subsequent words in a sentence from a probability distribution conditioned by the preceding text. However, our findings demonstrate that LLMs can capture aspects of human behavior extending beyond mere text production, which is the typical focus of standard benchmarks. In contrast to assessments based on single-response evaluations, our approach involves comparing the entire distribution of responses. We show that in analogous scenarios, humans and generative agents exhibit similar linking probability distributions. This observation is crucial for ABMs, which rely on Monte Carlo sampling and necessitate multiple updates of each agent to explore the full range of possible responses. Therefore, for successful integration of LLMs into ABMs, it is imperative to ensure that these models not only replicate realistic human-like responses in isolated instances but also accurately represent the entire probability distribution of potential human behaviors in a situation in which agents interact with each other.

\end{document}